\documentclass{Interspeech2024}

\interspeechcameraready

\title{Automatic detection of Mild Cognitive Impairment using high-dimensional acoustic features in spontaneous speech}

\name[affiliation={1}]{Cong}{Zhang}
\name[affiliation={2}]{Wenxing}{Guo}
\name[affiliation={1}]{Hongsheng}{Dai}

\address{
  $^1$Newcastle University, UK\\
  $^2$Essex University, UK}
\email{cong.zhang@newcastle.ac.uk, wg22745@essex.ac.uk, hongsheng.dai@newcastle.ac.uk}


\begin{document}

\maketitle

\begin{abstract}
    This study addresses the TAUKADIAL challenge, focusing on the classification of speech from people with Mild Cognitive Impairment (MCI) and neurotypical controls. We conducted three experiments comparing five machine-learning methods: Random Forests, Sparse Logistic Regression, k-Nearest Neighbors, Sparse Support Vector Machine, and Decision Tree, utilizing 1076 acoustic features automatically extracted using openSMILE. In Experiment 1, the entire dataset was used to train a language-agnostic model. Experiment 2 introduced a language detection step, leading to separate model training for each language. Experiment 3 further enhanced the language-agnostic model from Experiment 1, with a specific focus on evaluating the robustness of the models using out-of-sample test data. Across all three experiments, results consistently favored models capable of handling high-dimensional data, such as Random Forest and Sparse Logistic Regression, in classifying speech from MCI and controls.
\end{abstract}

\section{Introduction}
\subsection{Mild cognitive Impairment}
With ageing emerging as an increasingly prevalent global trend, cognitive impairment has become an ever more concerning topic. The scarcity of medical resources poses significant challenges for the timely detection of such cognitive impairment as dementia. The early stage of dementia is commonly identified as mild cognitive impairment (MCI). MCI often, though not necessarily, progresses into full-blown dementia. People with MCI may self-report subjective cognitive difficulties, e.g. forgetting recent events, yet their speech typically does not exhibit obvious distinctions from peers of a similar age. Given the neurodegenerative nature of these conditions, early intervention is crucial. However, the subtle nature of early-stage symptoms makes accurate diagnosis challenging. Therefore, automatic assessment emerges as a valuable tool, offering easy accessibility for people with MCI and a convenient reference for clinicians. 

The review in \cite{muellerConnectedSpeechLanguage2018} presents thoroughly the distinction between MCI and typical speech, summarising the effects of semantics, syntactic complexity, fluency, vocal parameters, and pragmatics. For speech-related factors, MCI speakers exhibit less fluent speech patterns, different intonation for complex descriptions, and different speech rates and phonation patterns.

\subsection{Automatic classification}
Compared with the classification of dementia (e.g. \cite{iliasMultimodalDeepLearning2022,chenAutomaticDetectionAlzheimer2021}), relatively early-stage MCI has received less attention. \cite{gosztolyaIdentifyingMildCognitive2019} trained classification models for MCI and Alzheimer's disease with acoustic features, including Automatic Speech Recognition outputs (i.e. text transcriptions with filled and unfilled pauses) and other linguistic features such as morphological or semantic features. The combination of both acoustic and other linguistic features outperformed either unimodal input. They also compared acoustic features extracted based on manual and automatic segmentation and found that manual segmentation had higher accuracy. In \cite{nagumoAutomaticDetectionCognitive2020}, temporal features (such as duration of utterance, number and length of pauses) and spectral features (F0, F1, and F2) were used to build logistic regression models to predict MCI, cognitive impairment, MCI with global cognitive impairment and the controls. The MCI classification accuracy, assessed through a 3-fold cross-validation, yielded a result of 0.61. More recently, as a part of the TAUKADIAL Challenge, \cite{LuzEtAlTAUKADIAL24} tested \texttt{eGeMAPs} features (88 features), a self-supervised feature learning approach, \texttt{wav2vec} (512 features), and the two combined. After a 20-fold cross-validation, the highest performing model,  \texttt{wav2vec + eGeMAPs},  presented a UAR of 59.18.

\subsection{The current study}
This paper is submitted as a part of the TAUKADIAL Challenge, focusing on creating models to classify speech from people with MCI and controls. To achieve this goal, we set out to investigate: (1) what acoustic features are needed to successfully predict the classification; (2) what modelling technique can best classify the data; (3) whether a fully automated process, i.e. without any manual annotation and labelling, can lead to an effective classification.

\section{Methods}
\subsection{Data}
The data used in this study was distributed as `training data' by the TAUKADIAL organizers\footnote{Test set details have not been published thus not included.}. The full dataset is described in \cite{LuzEtAlTAUKADIAL24}. This subset of data contain 387 files from 129 distinct speakers (three files per speaker), including English data (American English) and Chinese data (Taiwan Mandarin). The language variety information was not given explicitly.

\subsection{Language identfication}
The Python package \texttt{VoxLingua107} \cite{valk2021slt} was used for language identification. The English data, being clearer and louder, resulted in mostly accurate detection; the Mandarin data included speakers speaking with various accents, and therefore the detection results were not ideal. Consequently, we recorded the identification results for English and classified the rest as Mandarin. Out of 383 files (129 speakers), 182 files (62 speakers) were identified as English, while the remaining 201 files (67 speakers) were categorized as Mandarin.

\subsection{Feature extraction}
We extracted acoustic features from the raw audio files employing the openSMILE Toolkit \cite{eyben2010opensmile}, which was specifically developed for audio feature extraction and classification of speech and music signals. Two feature sets, \texttt{emobase} (988 features) and \texttt{eGeMAPSv02} (88 features), were extracted using the Python package \texttt{opensmile}\footnote{https://github.com/audeering/opensmile-python} in Python 3.9. These two feature sets provide sufficient coverage of consonant and vowel features, voice quality features, and prosodic features, which are all relevant to MCI speech. We did not process these data any further to keep the process as automated as possible.

\subsection{Feature selection}
We employed regularization methods for feature selection, which is a popular machine-learning technique aimed at reducing overfitting and improving the generalization ability of models by adjusting the weights of features. This typically involves introducing a regularization term into the loss function, which is a function of the feature weights, to constrain the magnitude of feature weights or induce sparsity. See more details in the description of the SLR and SSVM methods in Section 2.5.

\subsection{Classification methods}
Considering the high-dimensional nature of the dataset, where the number of features exceeds the number of samples and significant correlations exist among the data, we employed five model training methods: Random Forests (RF), Sparse Logistic Regression (SLR), k-Nearest Neighbors (KNN), Sparse Support Vector Machine (SSVM), and Decision Tree (DT).

Random Forests is a classifier comprised of a collection of tree-structured classifiers \cite{breiman2001}. 
It constructs a multitude of decision trees on randomly selected sub-samples of the original dataset and aggregates the results from each tree to make predictions. During the construction of each tree, a random subset of features is considered for node splitting to increase model diversity. The prediction process involves employing a voting mechanism for classification tasks.
For classification, the predicted class \( \hat{y} \) is determined by:
\[ \hat{y} = \text{arg~max}_{k} \sum_{i=1}^n I(\hat{y}_i = k), \]
where \( \hat{y}_i \) represents the predicted class by the $i$ th tree, \( n \) is the total number of trees, \( k \) is the class label, and \( I(\cdot) \) is the indicator function.

Sparse Logistic Regression is a method that combines Elastic Net regularization with logistic regression \cite{jeromefriedman2010}.
The method uses both $L_1$ regularization (Lasso) and $L_2$ regularization (Ridge) to overcome some limitations of each individual method. In logistic regression, the objective function of Elastic Net is typically defined as:
\begin{align*}
&-\frac{1}{n} \sum_{i=1}^n \left[ y_i \log(p_i) + (1 - y_i) \log(1 - p_i) \right] \\
&+ \lambda_1 ||\beta||_1 + \lambda_2 ||\beta||_2^2, 
\end{align*}
where \( n \) is the number of samples, \( y_i \) is the observed response (class labels), \( p_i \) is the predicted probability for each observation, \( \lambda_1 \) and \( \lambda_2 \) are hyperparameters controlling the strength of $L_1$ and $L_2$ regularization penalties, \( ||\beta||_1 \) is the $L_1$ norm (sum of absolute values), \( ||\beta||_2^2 \) is the squared $L_2$ norm (sum of squared parameters), and \( \beta \) is the coefficient vector of the model.
By tuning the values of \( \lambda_1 \) and \( \lambda_2 \), one can balance the model's fit and complexity, leading to improved generalization performance. The main advantage of Elastic Net is its ability to handle high-dimensional data and correlated features, while selectively retaining some relevant features and compressing others to zero.

Sparse Support Vector Machine is an extension of the traditional Support Vector Machine algorithm aimed at inducing sparsity in the solution \cite{congruiyi2017}. The objective function of SSVM is typically defined as:

\[
\frac{1}{n} \sum_{i=1}^n \text{hingeLoss}\left(y_i\left(x_i^{\prime} w+b\right)\right) + \text{penalty}(w),
\]
where $x_i$ represents the feature vector of the $i$ th sample, $y_i$ denotes the true label of the $i$ th sample, $w$ is the weight vector of the model, $b$ stands for the bias term of the model and hingeLoss $(t)=$ $\max (0,1-t)$ is the hinge loss function. For this method, we still applied the Elastic Net penalty. Namely, $\text{penalty}(w)=\lambda_1 ||w||_1 + \lambda_2 ||w||_2^2$.
SSVM aims to find a hyperplane that separates the classes while using only a subset of the features, effectively reducing the dimensionality of the problem and improving computational efficiency. 

k-Nearest Neighbors classification method is a non-parametric and instance-based learning algorithm used for classification tasks. It operates by classifying an unseen data point based on the majority class among its $k$ nearest neighbors in the feature space. The choice of $k$, the number of neighbors to consider, is a crucial hyperparameter that influences the model's performance. Larger values of $k$ result in smoother decision boundaries but may lead to less flexibility, while smaller values of $k$ provide more flexible decision boundaries but may lead to overfitting. 

The Decision Tree algorithm is a popular machine learning method used for both classification and regression tasks. It builds a tree-like structure by recursively splitting the dataset into subsets based on the feature that provides the most significant information gain or decrease in impurity. Each internal node of the tree represents a decision based on a feature, while each leaf node corresponds to a class label or a regression value. Decision trees are interpretable and easy to visualize, making them useful for understanding the decision-making process.

\subsection{Evaluation metrics}
We used the following metrics to evaluate the $\mathrm{MCI}$ classification task.
Specificity ($\sigma$):
$
\sigma=\frac{T_N}{T_N+F_P};
$
Precision ($\pi$):
$
\pi=\frac{T_P}{T_P+F_P};
$
$F_1$ score:
$
F_1=\frac{2 \pi \rho}{\pi+\rho};
$
Balanced accuracy (unweighted average recall, UAR):
$
\mathrm{UAR}=\frac{\sigma+\rho}{2},
$
where $\rho=\frac{T_P}{T_P+F_N}$ denotes sensitivity that measures the proportion of actual positive cases, 
$N$ is the number of patients, $T_{P}$ represents the number of true positives, $T_{N}$ denotes the number of true negatives, $F_{P}$ is the number of false positives and $F_{N}$ stands for the number of false negatives. These metrics collectively provide a comprehensive assessment of the performance of $\mathrm{MCI}$ classification.

\section{Experiments}
In Experiment 1, the models were trained with all data, including English and Chinese, and tested on an in-sample test set. Experiment 2 included a language detection step and the models were trained separately. Experiment 3 further enhanced the language-agnostic model in Experiment 1, but tested the robustness of the models using out-of-sample test data. All models were trained and tested in R \cite{rcoreteam}. Some relevant R packages were used to compare these methods: Random Forests (R package \texttt{ranger}\cite{Wright2017ranger} with parameters: num.trees = 500, mtry=the square root of the number of variables), Sparse Logistic Regression (R package \texttt{glmnet}\cite{glmnet} with parameters: alpha=0.5, family = ``binomial", s = 0.001, type = ``response"), k-Nearest Neighbors (R package \texttt{class} \cite{class} with parameter: k = 5), Sparse Support Vector Machine (R package \texttt{sparseSVM} \cite{sparseSVM} with parameter: lambda = 0.01), Decision Tree (R package \texttt{rpart} \cite{rpart} with parameter: method = ``class"). Data and scripts are available\footnote{Anonymous link for peer review: https://tinyurl.com/n3nkepzr}.

\subsection{Experiment 1}
In Experiment 1, all available data were used for model training. A test set of 100 samples (out of 387) were randomly selected from the same training dataset for evaluating the trained models. 

The results in Table \ref{training data from English and Chinese speech} demonstrate that RF and SLR have exceptional performance with perfect precision, specificity, F1 score and UAR of $100\%$, followed by DT with high precision, specificity, F1 score, and UAR of $98\%$, $97\%$, $94\%$, and $94\%$ respectively. SSVM exhibits moderate performance with precision, specificity, F1 score, and UAR of $84\%$, $75\%$, $90\%$, and $86\%$ respectively, while KNN shows comparatively lower performance across all metrics.

\begin{table}[h]
\centering
\caption{Comparison of different methods used for training data from English and Chinese speech}
\label{training data from English and Chinese speech}
\begin{tabular}{ccccc}
\hline Method & Precision & Specificity & F1 score & UAR \\
\hline RF & $ 100 \%$ & $100 \%$ & $100 \%$ & $100 \%$ \\
\hline DT & $98 \%$ & $97 \%$ & $94 \%$ & $94 \%$ \\
\hline KNN & $77 \%$ & $68 \%$ & $78 \%$ & $73 \%$ \\
\hline SLR & $100 \%$ & $100 \%$ & $100 \%$ & $100 \%$ \\
\hline SSVM & $84 \%$ & $75 \%$ & $90 \%$ & $86 \%$ \\
\hline
\end{tabular}
100 samples are randomly selected from the same training dataset for evaluating the trained models.
\end{table}

\subsection{Experiment 2}
In Experiment 2, we conducted separate analyses for English and Chinese data using the predicted language labels. Test sets of 100 samples (out of 201) from the Chinese data and 100 samples (out of 186) were randomly selected from the same training data for model evaluation. 

The results are shown in Tables \ref{training data from Chinese speech} and \ref{training data from English speech}. From these two tables, we can observe that RF, SLR and SSVM perform the best among the methods considered, while KNN consistently exhibits the poorest performance among all the methods used. The results from these language-specific models showed consistently high accuracy in the classification task as the language-agnostic models, which suggests that (1) language-agnostic models are powerful enough to make the classifications; and (2) smaller training data size did not influence the robustness of the models, especially the best-performing ones. 

\begin{table}[h]
\centering
\caption{Comparison of different methods used for training data from Chinese speech}
\label{training data from Chinese speech}
\begin{tabular}{ccccc}
\hline Method & Precision & Specificity & F1 score & UAR \\
\hline RF & $100  \%$ & $100 \%$ & $100 \%$ & $100 \%$ \\
\hline DT & $87 \%$ & $86 \%$ & $92 \%$ & $91 \%$ \\
\hline KNN & $77 \%$ & $82 \%$ & $68 \%$ & $71 \%$ \\
\hline SLR & $100 \%$ & $100 \%$ & $100 \%$ & $100 \%$ \\
\hline SSVM & $99 \%$ & $99 \%$ & $98 \%$ & $98 \%$ \\
\hline
\end{tabular}
100 samples are randomly selected from the same training dataset for evaluating the trained models.
\end{table}

\begin{table}[h]
\centering
\caption{Comparison of different methods used for training data from English speech}
\label{training data from English speech}
\begin{tabular}{ccccc}
\hline Method & Precision & Specificity & F1 score & UAR \\
\hline RF & $100  \%$ & $100 \%$ & $100 \%$ & $100 \%$ \\
\hline DT & $93 \%$ & $87 \%$ & $92 \%$ & $89 \%$ \\
\hline KNN & $75 \%$ & $41 \%$ & $82 \%$ & $66 \%$ \\
\hline SLR & $100 \%$ & $100 \%$ & $100 \%$ & $100 \%$ \\
\hline SSVM & $100 \%$ & $100 \%$ & $100 \%$ & $100 \%$ \\
\hline
\end{tabular}
100 samples are randomly selected from the same training dataset for evaluating the trained models.
\end{table}

\subsection{Experiment 3}
In Experiment 3, we further evaluated the predictive performance of these methods. The entire dataset was partitioned into training and test sets, using train-test ratios of 1:1 (i.e. 1 part allocated for training and 1 part for testing), 2:1, 3:1, 4:1 and 5:1 respectively. To enhance the stability of the models, the original dataset was randomly partitioned 100 times for each train-test split, and we conducted 100 rounds of training and testing for each method using random cross-validation.

\begin{figure}[h]
\centering  
 \includegraphics[width=0.45\textwidth]{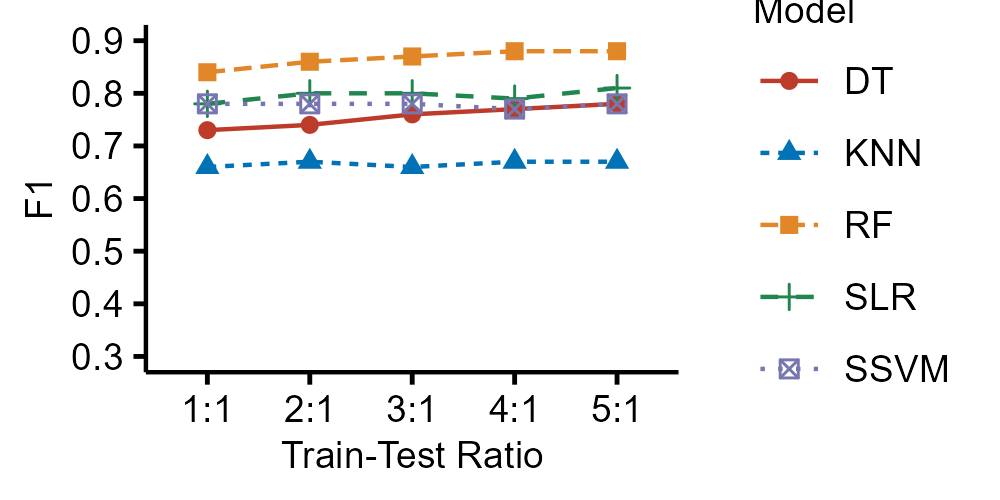}
\caption{F1 based on data split at random for 100 times. Train-test ratio denotes the proportions of the training and test sets.}  
\label{F1}
\end{figure}
\begin{figure}[h]  
\centering  
 \includegraphics[width=0.45\textwidth]{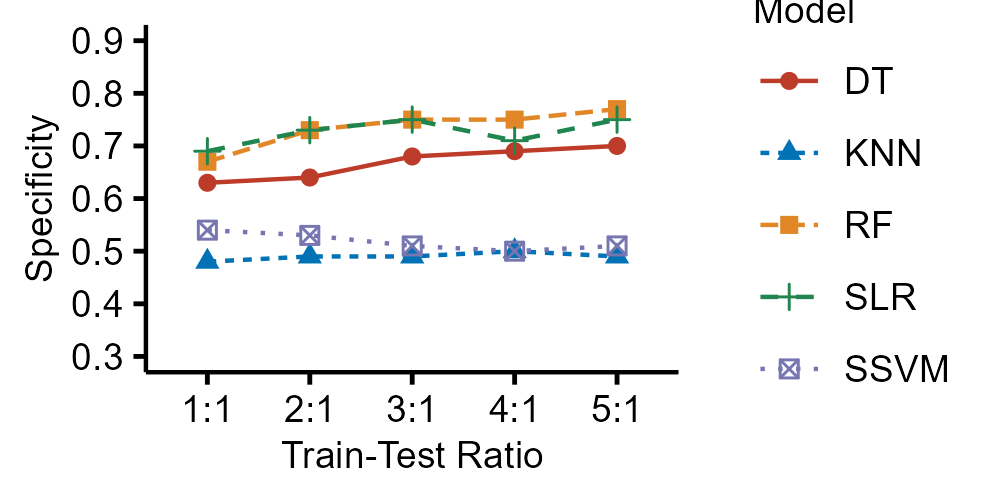}
\caption{Specificity based on data split at random for 100 times. Train-test ratio denotes the proportions of the training and test sets.}  
\label{Specificity}
\end{figure}
\begin{figure}[h]
\centering  
 \includegraphics[width=0.45\textwidth]{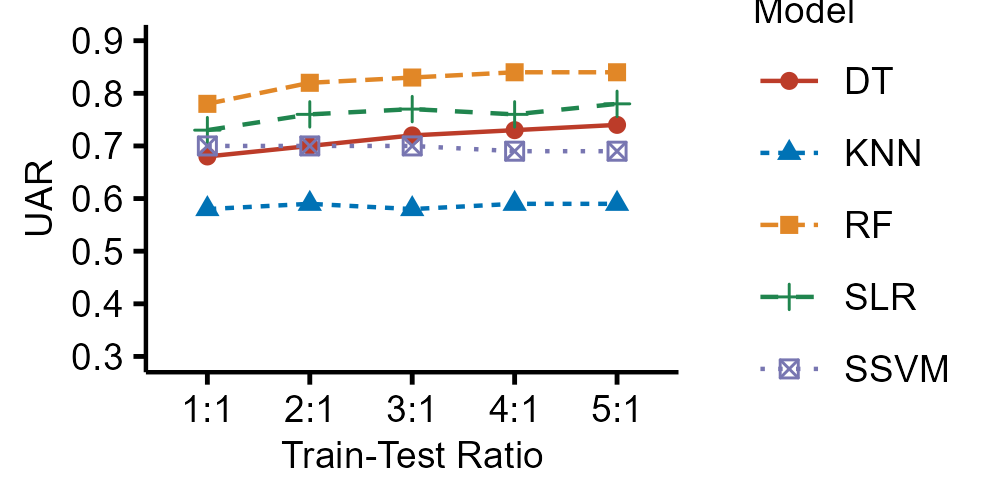}
\caption{Unweighted Average Recal based on data split at random for 100 times. Train-test ratio denotes the proportions of the training and test sets.}  
\label{UAR}
\end{figure}

Figures \ref{F1}, \ref{Specificity} and \ref{UAR} present the results of this experiment. From these three line graphs, we can see that RF consistently outperforms the other methods, exhibiting the highest scores across all three metrics. Specifically, in terms of F1 Score (Fig. \ref{F1}), RF achieves scores ranging from 0.84 to 0.88, followed by DT with scores ranging from 0.73 to 0.78; and then SLR and SSVM with scores ranging from 0.78 to 0.81. KNN consistently shows lower F1 scores compared to other methods, ranging from 0.66 to 0.67. 

Similarly, in terms of Specificity (Fig. \ref{Specificity}), RF demonstrates the highest values, ranging from 0.67 to 0.77, followed by SLR and DT with values ranging from 0.63 to 0.75 and 0.68 to 0.70, respectively. KNN consistently exhibits the lowest Specificity, ranging from 0.48 to 0.50.

Regarding UAR (Fig. \ref{UAR}), Random Forest again leads with scores ranging from 0.78 to 0.84, followed by Sparse Logistic Regression and Decision Tree with scores ranging from 0.68 to 0.78 and 0.72 to 0.74, respectively. K-Nearest Neighbors shows the lowest UAR, ranging from 0.58 to 0.59. These results suggest that Random Forest is the most effective method across all three metrics, followed by Decision Tree and Sparse Logistic Regression/Support Vector Machine, while K-Nearest Neighbors consistently performs the poorest.

Additionally, from these three graphs, we observed that the performance of the five methods based on the F1 score is superior to that of UAR and Specificity. These methods exhibit relatively stable performance in terms of F1 and UAR across different ratios of training and test datasets. Moreover, the performance of the five methods, except for SSVM, slightly improves with the increase in the ratio of training to testing datasets.

\section{Discussion and conclusions}
\subsection{Classification models}
By comparing the results across various methods in the three experiments, we observed that both the Random Forests method and Sparse Logistic Regression consistently demonstrate superior performance. This can be attributed to the inherent characteristics of these two methods, as follows. 

The Random Forests method typically achieves high prediction accuracy, particularly in handling large datasets and high-dimensional data. Moreover, it demonstrates robust resistance to overfitting by introducing randomness during tree construction, ensuring reliable generalization. Furthermore, the method exhibits robustness, as it can effectively utilize combinations of available samples and features for prediction.

Regarding logistic regression, we utilized Elastic Net penalties to induce sparsity. Logistic Regression with Elastic Net penalties offers several advantages. By combining both $L_1$ and $L_2$ penalties, Elastic Net regularization allows for simultaneous variable selection and regularization, addressing multicollinearity and reducing model complexity. This approach enhances robustness to overfitting by penalizing large coefficients, striking a balance between shrinkage and stabilization. Moreover, elastic net regression is particularly effective in handling high-dimensional datasets, where the number of predictors exceeds the number of observations. 

One possible reason for the comparatively lower performance of the k-Nearest Neighbors method across all metrics could be its sensitivity to noisy or high-dimensional data. KNN's prediction accuracy heavily relies on the local structure of the data, making it susceptible to noise and outliers. Additionally, the curse of dimensionality may impact KNN's performance negatively, as the distance between samples becomes less meaningful in high-dimensional spaces. 

\subsection{Implications and future studies}
The high accuracy of the best-performing models in all three experiments suggests that the number of features extracted can sufficiently account for the data. These features are purely based on automatic extraction from speech data therefore suggesting the possibility of developing a fully automatic speech-based tool for easy classification. With such a tool, clinicians will have a useful assistive tool to refer to when making clinical decisions. They can also use this tool to monitor patient progress before and after interventions. The results from the models also informs other studies of what types of models are best at dealing with high-dimensional data, e.g. \cite{zhang2024prosody}.

While the results have been good, we acknowledge that there are still many future directions to pursue. First, we would like to follow up the study with different cross-validation methods, such as Leave-One-Subject-Out cross-validation, to further test the predictive power of the models. Another direction for future research is to combine other features that are readily available, e.g. demographic information, and cognitive test scores; or can be automatically extracted, e.g. text-based features. Lastly, although the current features are sufficient in modelling the current data, we will continue to consider other acoustic features such as time-series acoustic features. Such features can include acoustic features that are not present in the current feature set, such as the number and length of pauses.

\subsection{Conclusions}
To conclude, it is possible to have a solely speech-based automatic detection tool for MCI speech. The features extracted using the \texttt{emobase} and \texttt{eGeMAPSv02}from \texttt{opensmile}\ can sufficiently account for the data. Models that can handle high-dimensional data, such as Random Forest and Sparse Logistic Regression, are strongly recommended in processing large sets of features as what have been used in the current study.



\bibliographystyle{IEEEtran}
\bibliography{template}

\end{document}